\documentclass[preprint,12pt]{elsarticle}

\usepackage{amssymb}
\usepackage{amsmath}
\usepackage{xcolor}

\journal{Communications in Nonlinear Science and Numerical Simulation}

\begin{document}

\begin{frontmatter}



\title{Noise-induced transitions between coexisting limit cycles beyond saddle-node bifurcations}


\author[urjc]{Ignacio Ortega-Piwonka\corref{email}} 
\author[urjc]{Javier Used}
\author[urjc]{Jes\'{u}s M. Seoane}
\author[urjc,rac]{Miguel A.F. Sanju\'{a}n}

\affiliation[urjc]{
organization={Nonlinear Dynamics, Chaos and Complex Systems Group, Universidad Rey Juan Carlos},
            addressline={Tulip\'{a}n s/n}, 
            city={M\'{o}stoles},
            postcode={28933},
            state={Madrid},
            country={Spain}}
            
\affiliation[rac]{organization={Royal Academy of Sciences of Spain},
            addressline={Valverde 22}, 
            city={Madrid},
            postcode={28004},
            state={Madrid},
            country={Spain}}

\cortext[email]{juanignacio.ortega@urjc.es}

\begin{abstract}
Nonlinear multistable systems with limit cycles may exhibit noise-induced transitions between oscillatory states, even in parameter regimes beyond the saddle-node bifurcations that determine the existence of these limit cycles. This behavior is associated with the ghost of a saddle-node bifurcation, a well-known feature of nonlinear systems near such bifurcation points. To investigate this phenomenon, we first consider a stochastic minimal model that supports two coexisting limit cycles within a parameter interval bounded by two saddle-node bifurcations. We numerically and theoretically estimate the probability of remaining near each stable attractor as a function of both the noise intensity and the distance to the ghost structure. The model is then extended by introducing spatially localized, non-homogeneous noise with enhanced intensity in a prescribed region of phase space. This modification is intended to emulate the asymmetric structure of the Hindmarsh--Rose model, in which bursting limit cycles approach each other closely in specific regions of phase space, thereby facilitating noise-induced transitions between them.
\end{abstract}

\begin{highlights}
\item A bi-variate, axisymmetric model with homogeneous noise is proposed and studied.
\item The birhythmicity exhibited by the model is extended by the noise.
\item The model is numerically and analytically characterized.
\item The model is amended by incorporating non-homogeneous noise.
\item The model reproduces behaviors observed in neurological nonlinear systems.
\end{highlights}


\end{frontmatter}



\section{Introduction} \label{sec_Introduction}

Multistability is a widespread phenomenon in nonlinear systems \cite{Daza_et_al_Physics_Today_2024}. It refers to the coexistence of multiple attractors, including fixed points, limit cycles, quasiperiodic regimes, and chaotic attractors, among others. The attractor eventually reached by the system depends on the initial conditions \cite{Strogatz, Izhikevich}. A particularly relevant instance of multistability is birhythmicity, which denotes the coexistence of two periodic solutions with distinct amplitudes and frequencies \cite{Slepukhina_et_al_Chaos_2023}.

Understanding the effects of noise on multistable systems is relevant across several areas of science, including physics, chemistry, and biology \cite{Bashkirtseva_2020, Shukrinov_2017, Arumugan_2017}. Noise plays a fundamental role in neuronal dynamics, both at the level of individual neurons and in the collective behavior of neuronal networks \cite{Used_2024, Zirkle_2021}. When noise is introduced into a multistable system, random transitions between coexisting attractors may occur, as stochastic fluctuations can drive trajectories across the separatrices that delimit their basins of attraction \cite{Slepukhina_et_al_Chaos_2023}. For example, noise-induced chaos beyond the deterministic region of existence of a chaotic attractor has been reported in a neuronal discrete map \cite{Ryashko_Bashkirtseva_CSF_2023}. In epilepsy research, noise has also been shown to trigger transitions that lead to epileptic seizures \cite{Shen_et_al_CFS_2023}.

Multistable systems exhibiting one or more stable limit cycles in parameter regimes structured by saddle-node bifurcations are of particular interest, since transitions associated with these cycles can occur even beyond the corresponding bifurcation point. This behavior is associated with the ghost of a saddle-node bifurcation, a characteristic feature of nonlinear systems near these bifurcations \cite{Izhikevich}. Beyond the bifurcation, neither the stable nor the unstable limit cycle exists. However, if the initial condition lies close to the region of phase space where the annihilated limit cycles used to exist, the trajectory may remain in this region for a finite time before escaping from it. Consequently, in the presence of noise, transitions mediated by this ghost transient may occur and are qualitatively similar to the stochastic transitions observed inside the coexistence region, where an actual stable limit-cycle attractor is present.

The Hindmarsh--Rose model provides a relevant framework for investigating the effects of noise on birhythmicity. It was proposed in 1984 to reproduce phenomena commonly observed in experiments on single neurons, such as bursting, slow-fast dynamics, and diverse firing patterns \cite{Hindmarsh_Rose_ProcRoySoc_1984, Barrio_Chaos_2020_HR_homoclinic, Slepukhina_et_al_Chaos_2023}. As the parameter associated with the fast ion channels is varied, the model exhibits a branch of bursting limit cycles that undergoes successive folds through a sequence of saddle-node bifurcations, with an increasing number of spikes per burst. Regions of birhythmicity arise between consecutive bifurcations. When noise is introduced, transitions occur between the coexisting bursting limit cycles. Owing to the ghost phenomenon, such transitions can also occur outside, but close to, the birhythmicity regions. In Ref.~\cite{Ortega_et_al_PRE_2025}, the phenomenon of stochastic birhythmicity in the Hindmarsh--Rose model was investigated in detail, and a minimal model was proposed to provide a qualitative description of the underlying mechanism. However, that model captures the phenomenon only partially, as it exhibits a single limit cycle coexisting with a fixed-point attractor rather than two coexisting limit cycles, and a single saddle-node bifurcation of limit cycles rather than a pair of such bifurcations. Consequently, it does not display genuine birhythmicity, although it does exhibit the noise-induced emergence of periodic oscillations beyond the saddle-node bifurcation.

To gain a deeper understanding of the mechanisms underlying birhythmicity in systems with a structure similar to that of the Hindmarsh--Rose model, as well as of the role played by the ghost phenomenon, we propose here a stochastic, axially symmetric birhythmic model in which two coexisting limit cycles arise within a parameter interval bounded by two saddle-node bifurcations. The simplicity of the model allows for a tractable probabilistic description based on the associated Fokker--Planck equation \cite{Gardiner}. Within this framework, stochastic transitions between the coexisting limit cycles may occur throughout the phase space, in contrast to the strongly asymmetric Hindmarsh--Rose model, where such transitions are largely confined to regions in which the limit-cycle trajectories approach each other closely.

To capture this effect, we extend the simplified model by introducing non-homogeneous noise with enhanced intensity in a prescribed region of phase space. As a consequence, stochastic transitions become more likely to occur within that region. Overall, both the homogeneous- and non-homogeneous-noise models qualitatively reproduce the same hourglass-like structure observed in the parameter space of the Hindmarsh--Rose model, suggesting that this structure may be a robust feature of nonlinear systems in which two limit-cycle attractors coexist within a region bounded by saddle-node bifurcations. This hourglass-like structure emerges from long-time observations throughout parameter space. However, in the low-noise regime, even very long observation times may be insufficient for the system to reach stochastic equilibrium, since the corresponding relaxation times become exceedingly large and practically inaccessible. Consequently, a theoretical approach based on stationary probability distributions is proposed to describe the behavior of the system when the noise is high, while in the low-noise case, a complementary approach, inspired by Kramers' escape problem \cite{Kramers_Physica_1940, Chandrasekhar_1943}, is proposed.

The article is organized as follows. In Sec.~\ref{sec_simple_model}, we introduce the axially symmetric model and characterize stochastic birhythmicity in terms of the noise intensity and a control parameter. In Sec.~\ref{sec_simple_model_LN}, we incorporate non-homogeneous noise into the model and analyze the resulting stochastic birhythmicity. Finally, Sec.~\ref{sec_conclusions} summarizes the main conclusions.

\section{Double saddle-node bifurcation model with homogeneous noise}\label{sec_simple_model}

\subsection{Deterministic model}

We introduce a simple model designed to emulate the birhythmicity observed in the Hindmarsh--Rose model, where the coexisting attractors remain relatively far from the origin but come into close proximity in a specific region of phase space. This model exhibits two coexisting stable limit cycles bounded by two saddle-node bifurcations. The model is a bivariate system of stochastic differential equations written in vector form as
\begin{equation}\label{eq_simple_model}
\frac{d\mathbf{r}}{dt} = \mathbf{F}(\mathbf{r}) + \varepsilon\boldsymbol{\xi}(t),
\end{equation}
where $\mathbf{r}=\big(x(t),y(t)\big)$ collects the two time-dependent variables. The term $\mathbf{F}(\mathbf{r})$ represents the deterministic dynamics, while $\varepsilon \boldsymbol{\xi}(t)$ accounts for intrinsic noise. In biological neurons, such noise reflects the intrinsic complexity of the system and the large number of underlying components and variables involved. The parameter $\varepsilon$ controls the noise intensity, and $\boldsymbol{\xi}(t)$ is a bivariate, uncorrelated white-noise process \cite{Gardiner}, defined by
\begin{align}\label{eq_xi}
  \begin{split}
    \langle \xi_i(t)          \rangle &= 0, \\
    \langle \xi_i(t) \xi_j(t')\rangle &= \delta_{ij}\delta(t'-t),
  \end{split}
\end{align}
where $\delta_{ij}$ represents the Kronecker delta coefficients and $\delta(t'-t)$ is the Dirac delta function. The deterministic component is given by an autonomous vector field, which in polar coordinates naturally separates into radial and angular parts,
\begin{equation}\label{eq_F}
\mathbf{F}(\mathbf{r}) = f(r)\mathbf{\hat{r}} + r\omega\boldsymbol{\hat\theta},
\end{equation}
where $r=|\mathbf{r}|$ is the distance to the origin, while $\mathbf{\hat{r}}$ and $\boldsymbol{\hat\theta}$ are the polar unit vectors. The parameter $\omega$ represents the angular velocity around the origin, which throughout this section is fixed at $\omega=1$. This choice is arbitrary and does not affect the long-time behavior of the system. The radial component is defined by the scalar function $f(r)$,
\begin{equation}\label{eq_f}
f(r) = -r \big( N(r^2)-b \big),
\end{equation}
where $b$ is a control parameter and $N(r^2)$ is a scalar function with an N-shaped profile,
\begin{equation}\label{eq_N}
N(r^2) = \frac{10}{r^2}(r^2-2.25^2)(r^2-2.50^2)(r^2-2.75^2).
\end{equation}
The specific values $2.25$, $2.50$, and $2.75$ used in Eq.~(\ref{eq_N}) are chosen so that the distance between the two coexisting limit cycles remains relatively small compared to their radii. We have also tested other choices of these values, as well as different N-shaped profiles, and verified that no significant qualitative changes arise in the results presented in this work.

The function $N(r^{2})$ vanishes at $r=2.25$, $r=2.50$, and $r=2.75$, diverges to $-\infty$ as $r \to 0$, and diverges to $\infty$ as $r \to \infty$. It possesses a local maximum at $(r,N)=(2.354,1.238)$ and a local minimum at $(r,N)=(2.643,-1.168)$.

\begin{figure}
\centering\includegraphics[width=0.5\linewidth]{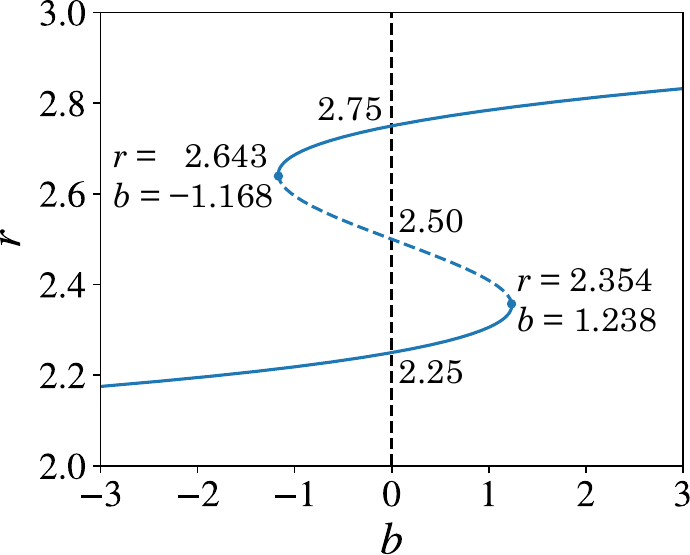}
\caption{Bifurcation diagram of the deterministic simple model, showing the limit cycle radius $r$ as a function of the control parameter $b$. The solid blue lines denote the stable limit cycle branches and dashed blue line denotes the unstable limit cycle branch. The diagram also marks the limit cycle saddle-node bifurcations, and the intersections of the branches with the $b = 0$ axis.}
\label{fig_DWP_det_bif_diag}
\end{figure}

The deterministic system defined by Eq.~(\ref{eq_simple_model}) with $\varepsilon=0$ has a singular point at $r=0$, where $f(r)$ diverges. This point is always repelling, regardless of the values of the parameters $b$ and $\omega$. The equilibrium solutions correspond to limit cycles centered at the origin and rotating with angular velocity $\omega$. Their radii are given by the nonzero solutions of $f(r)=0$, or equivalently by the solutions of $N(r^{2})=b$. Owing to the N-shaped profile of $N(r^{2})$, this equation admits a single non-negative solution for $b<-1.168$ or $b>1.238$, and three non-negative solutions for $-1.168<b<1.238$. Consequently, the system possesses lower, middle, and upper branches of limit-cycle solutions, as illustrated in Fig.~\ref{fig_DWP_det_bif_diag}. The lower and upper branches are stable, whereas the middle branch is unstable. Their stability is determined by the sign of the derivative of $f(r)$. The lower and middle branches coalesce in a saddle-node bifurcation of limit cycles at $(b,r)\approx(1.238,2.354)$, whereas the middle and upper branches coalesce in a second saddle-node bifurcation at $(b,r)\approx(-1.168,2.643)$. These bifurcations are indicated by dots in Fig.~\ref{fig_DWP_det_bif_diag}. Therefore, the system exhibits birhythmicity for $-1.168<b<1.238$. As $b\to-\infty$, the radius of the lower limit-cycle branch tends asymptotically to zero.

\begin{figure}
\centering
\includegraphics[width=\linewidth]{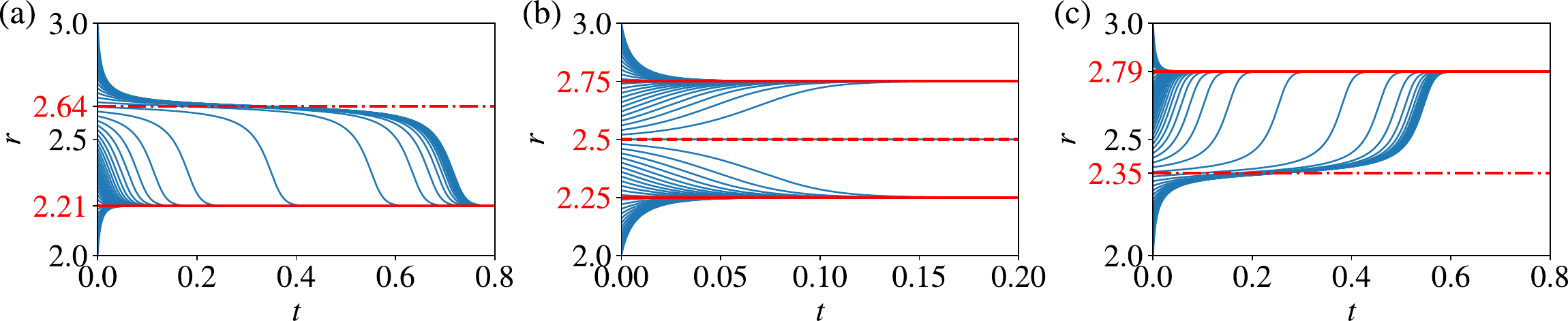}
\caption{Simulations of Eq.~(\ref{eq_simple_model}) for $\varepsilon=0$ and different values of $b$: (a) $b=-1.2$, (b) $b=0$, and (c) $b=1.3$. The initial conditions are uniformly distributed between $r=2$ and $r=3$. Time series of $r(t)$ are shown as solid blue curves. The radii of the stable and unstable limit cycles are indicated by solid and dashed red curves, respectively. In panels (a) and (c), the radius of the nearby saddle-node bifurcation is also shown as a dash-dotted red line. Due to the ghost phenomenon, $r(t)$ remains close to this radius for some time before approaching the stable limit cycle.}
\label{fig_DWP_ghosts}
\end{figure}

Some simulations of the deterministic system for different initial conditions and values of $b$ are shown in Fig.~\ref{fig_DWP_ghosts}, where the distance to the origin, $r(t)$, is plotted as a function of time. Within the birhythmicity region (Fig.~\ref{fig_DWP_ghosts}(b)), the system converges to one of the two stable limit cycles depending on whether the initial distance $r(0)$ is smaller or larger than the radius of the unstable limit cycle located between them. In other words, the unstable cycle acts as a separatrix between the basins of attraction of the two stable cycles. Outside the birhythmicity region, only a single stable limit cycle exists, and the system converges to it regardless of the initial condition. However, near the boundaries of the birhythmicity region, the ghost phenomenon emerges \cite{Izhikevich}. In Fig.~\ref{fig_DWP_ghosts}(a), $b$ is set slightly below $-1.168$, corresponding to the left saddle-node bifurcation shown in Fig.~\ref{fig_DWP_det_bif_diag}. In this case, the stable limit cycle has a radius of approximately $2.21$. If $r(0)<2.64$, corresponding to the radius of the colliding cycles at the bifurcation, the system converges rapidly to the stable limit cycle. By contrast, if $r(0)>2.64$, the trajectory slows down near $r=2.64$ and remains in its vicinity for a considerable time before eventually converging to the stable limit cycle. A similar effect is observed in Fig.~\ref{fig_DWP_ghosts}(c), where $b$ is set slightly above $1.238$, corresponding to the right saddle-node bifurcation in Fig.~\ref{fig_DWP_det_bif_diag}. In this case, the stable limit cycle has a radius of approximately $2.79$. If $r(0)>2.35$, corresponding to the radius at the saddle-node bifurcation, convergence is rapid. However, if $r(0)<2.35$, the trajectory slows down near $r=2.35$ before eventually converging to the stable limit cycle. Thus, the ghost phenomenon acts as a precursor of the stable limit cycle that disappears at each saddle-node bifurcation. This effect becomes increasingly pronounced as the system approaches the birhythmicity region: as $b$ approaches either saddle-node bifurcation, the time during which $r(t)$ remains near the corresponding saddle-node radius diverges.

\subsection{Stochastic model}

When noise is introduced into the model (i.e., $\varepsilon > 0$), the solutions $\mathbf{r}(t)$ of Eqs.~(\ref{eq_simple_model},\ref{eq_xi}) become stochastic processes. Their time evolution can be described statistically by a probability distribution $\rho(\mathbf{r},t)$, governed by the associated Fokker-Planck equation \cite{Gardiner, Fokker_FPeq_1914, Planck_FPeq_1917, Kolmogorov_FPeq_1931},
\begin{equation}\label{eq_FP}
\frac{\partial\rho}{\partial t} = -\nabla \cdot \big( \mathbf{F}(\mathbf{r}) \rho \big) + \frac{\varepsilon^2}{2} \nabla^2 \rho,
\end{equation}
where $\nabla \equiv (\partial/\partial x,\partial/\partial y)$ is the nabla operator. Equation~(\ref{eq_FP}) admits a stationary solution, namely a probability distribution to which the system converges after a sufficiently long time, independently of its initial state. Unlike the Hindmarsh--Rose model, which is strongly asymmetric, the present model is axially symmetric, i.e., statistically invariant under rotations. This symmetry, together with the existence of a stationary regime, ensures that the system is ergodic, so that expected values can be estimated as time averages \cite{Gardiner}. It is therefore natural to assume that the stationary distribution depends only on the radial distance to the origin, $\rho_{s}(\mathbf{r})=\rho_{s}(r)$. After removing the time dependence and using the identity $\nabla^2 \rho=\nabla\cdot(\nabla\rho)$, Eq.~(\ref{eq_FP}) yields
\begin{equation}\label{eq_FP_s0}
\nabla \cdot \Big( -\mathbf{F}(\mathbf{r}) \rho + \frac{\varepsilon^2}{2} \nabla \rho \Big) = 0.
\end{equation}
In polar coordinates, and provided that there is no dependence on $\theta$, it turns out that $\nabla \rho = \frac{\partial\rho}{\partial r} \mathbf{\hat{r}}$ and $\nabla\cdot\mathbf{H} = \frac{1}{r}\frac{\partial}{\partial r}(r H_r)$. Thus, Eq.~(\ref{eq_FP_s0}) yields
\begin{equation}\label{eq_FP_s1}
\frac{1}{r} \frac{d}{dr} \bigg( r \Big( -f(r)\rho_s + \frac{\varepsilon^2}{2} \frac{d\rho_s}{dr} \Big) \bigg) = 0.
\end{equation}
The radial component $f(r)$, defined in Eq.~(\ref{eq_f}), can also be written as the negative gradient of a scalar potential, $f(r) = -du/dr$, with
\begin{equation}\label{eq_u}
u(r) = \tfrac{1}{6} u_1 r^6 - \tfrac{1}{4} u_2 r^4 + \tfrac{1}{2} (u_3 - b) r^2 - u_4 \ln(r),
\end{equation}
where the coefficients take the values $u_{1} = 10$, $u_{2} = 188.75$, $u_{3} \approx 1171.91$, and $u_{4} \approx 2392.82$. By applying Eq.~(\ref{eq_u}) and imposing the normalization condition that the integral of $\rho_{s}$ over the entire plane equals unity, Eq.~(\ref{eq_FP_s1}) reduces to
\begin{equation}\label{eq_FP_s2}
\frac{du}{dr}\rho_s + \frac{\varepsilon^2}{2} \frac{d\rho_s}{dr} = 0.
\end{equation}
The solution of Eq.~(\ref{eq_FP_s2}) assuming zero stationary radial probability current is
\begin{equation}\label{eq_rho_s}
\rho_s(r) = Ce^{-2u(r)/\varepsilon^2},
\end{equation}
where $C$ is a normalization constant. This function gives the probability density of finding the system in a vector state $\mathbf{r}$ such that $|\mathbf{r}| = r$. The probability density of the scalar radial coordinate $r$ is obtained by integrating $\rho_{s}(r)$ along the circumference of radius $r$ centered at the origin,
\begin{equation}\label{eq_p_s}
p_s(r) = 2 \pi C r e^{-2u(r)/\varepsilon^2}.
\end{equation}

\begin{figure}
\centering
\includegraphics[width=\linewidth]{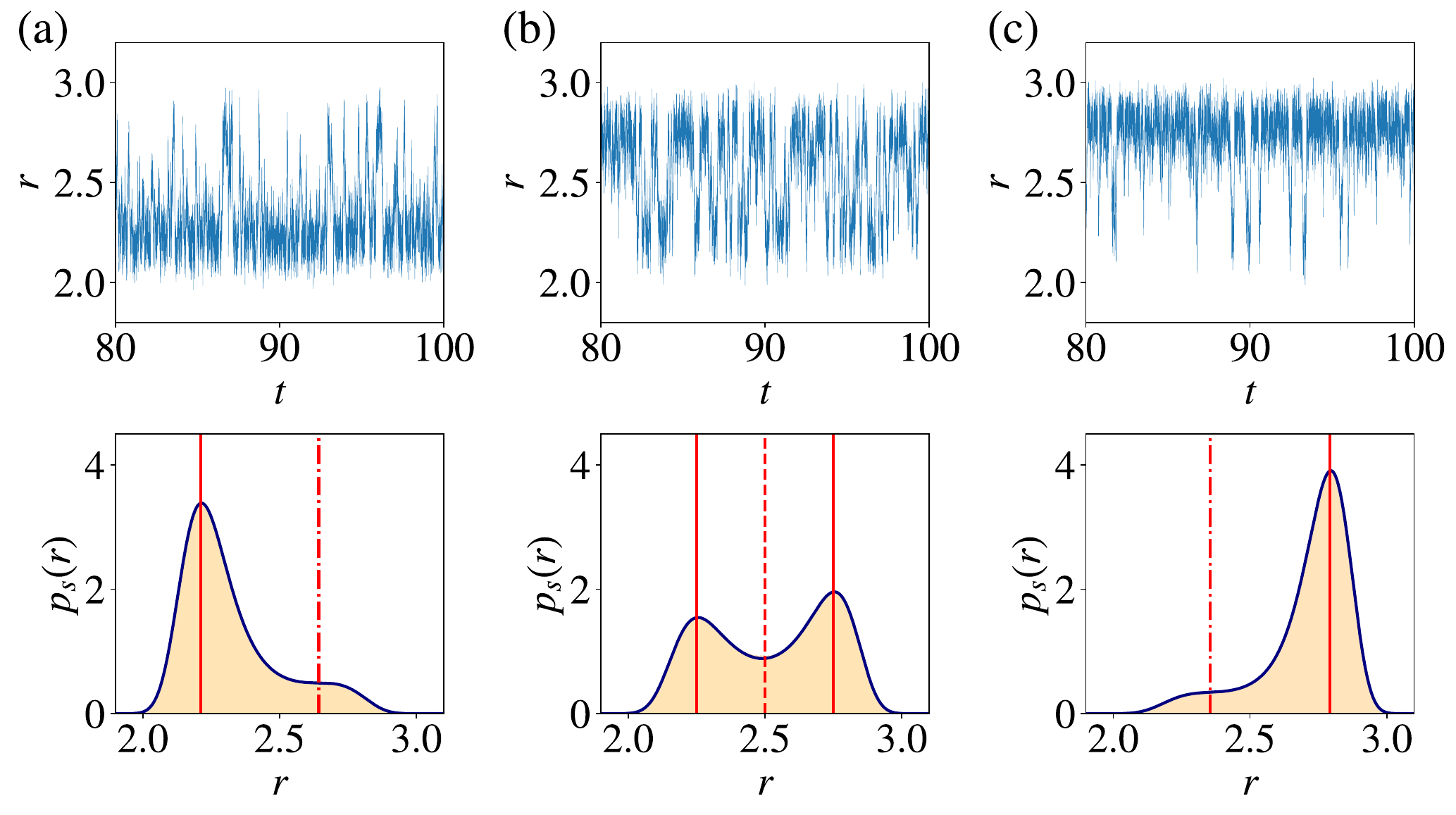}
\caption{Numerical simulations of Eqs.~(\ref{eq_simple_model},\ref{eq_xi}) for $\varepsilon=1.2$ and different values of $b$: (a) $b=-1.3$, (b) $b=0$, and (c) $b=1.3$. In each case, the time series $r(t)$ from a single simulation is shown after a transient, once the system has reached a steady state, together with the normalized histogram of $r$ obtained from 100 simulations of duration $\Delta t=100$ with initial conditions uniformly distributed between $r=2$ and $r=3$. Each histogram consists of 200 bins, shown in yellow, and is compared with the stationary probability distribution $p_s(r)$, shown as a solid blue curve. The radii of the stable and unstable deterministic limit cycles are indicated by solid and dashed red lines, respectively. In panels (a) and (c), the radius of the nearest saddle-node bifurcation is also shown as a dash-dotted red line, acting as a \textit{de facto} separatrix between the stable limit cycle and the ghost.}
\label{fig_DWP_samples}
\end{figure}

Simulations of Eqs.~(\ref{eq_simple_model},\ref{eq_xi}) for different parameter values $(b,\varepsilon)$ are shown in Fig.~\ref{fig_DWP_samples}, together with the corresponding histograms. Within the birhythmicity region (Fig.~\ref{fig_DWP_samples}(b)), bounded by the saddle-node bifurcations, the stochastic system undergoes random transitions between the two limit-cycle attractors. This behavior is reflected in the histogram of the radial distance $r$, which exhibits two peaks located at the radii of the deterministic stable cycles, separated by a minimum at the radius of the unstable cycle. Outside the birhythmicity region (Figs.~\ref{fig_DWP_samples}(a) and \ref{fig_DWP_samples}(c)), the system possesses a single stable limit-cycle attractor. When noise is introduced, the trajectory fluctuates around this attractor, but it may occasionally visit the ghost region associated with the annihilated cycle. This behavior is reflected in the corresponding histograms, which display a single peak centered near the radius of the deterministic attractor, together with a tail extending beyond the radius of the nearest saddle-node bifurcation.

This tail arises from the ghost phenomenon. Near a saddle-node bifurcation, noise may drive the system into the ghost region where the annihilated cycle would have existed, causing the trajectory to remain there for a relatively long time before eventually returning to the stable limit cycle. The numerically obtained histograms are in excellent agreement with the theoretically predicted stationary distribution $p_s(r)$. This agreement can be quantified by the Hellinger distance, which is of the order of $10^{-4}$ \cite{Gibbs_Su_2002}. As shown below, however, this agreement deteriorates in the low-noise regime because the system requires exceedingly long times to reach stochastic equilibrium.

Although, in the simple model studied here, the peaks and valleys of the state distributions coincide with the deterministic stable and unstable solutions, respectively, such a correspondence need not hold in other stochastic systems as the noise intensity increases \cite{Ortega_et_al_PRE_2010}. In the present case, however, this effect is barely noticeable, since the deterministic stable branches lie very close to one another within the birhythmicity region.

\begin{figure}
\centering
\includegraphics[width=\linewidth]{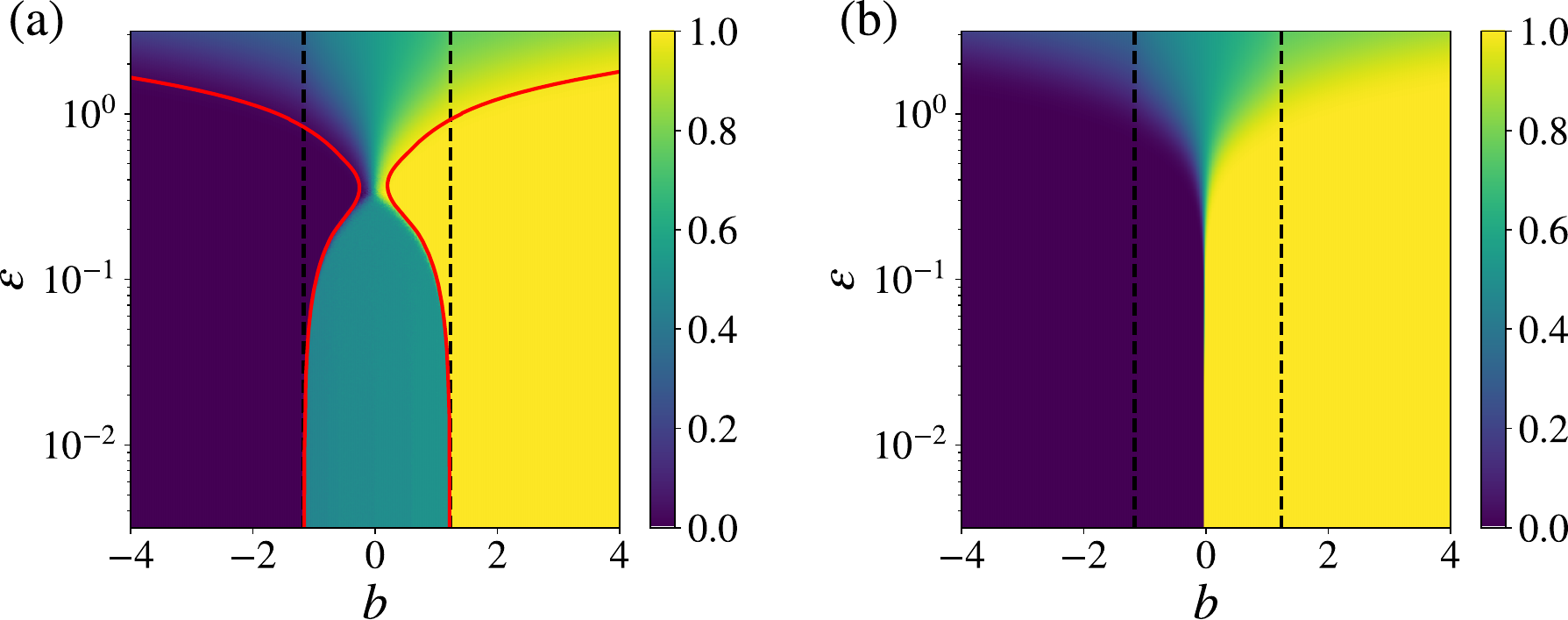}
\caption{(a) Numerical estimation of the probability $P_\text{LS}$ of observing the largest limit cycle, or its ghost, in the stochastic system, Eqs.~(\ref{eq_simple_model},\ref{eq_xi}), as a function of $(b,\varepsilon)$. For each parameter set, 100 simulations of duration $\Delta t=1000$ are performed, with initial conditions uniformly distributed between $r=2$ and $r=3$. After $t=500$, the fraction of time points for which $r(t)$ exceeds the radius of the separatrix is computed. The dashed black lines indicate the boundaries of the deterministic birhythmicity region, while the solid red lines mark the estimated boundaries of the \textit{de facto} stochastic birhythmicity region, defined as the region in which both states occur with probability at least $2\%$. (b) Numerical computation of the theoretical probability $P_{\mathrm{LS}}$, given by Eq.~(\ref{eq_P_LS}).}
\label{fig_DWP_param_space}
\end{figure}

Overall, Fig.~\ref{fig_DWP_samples} shows that, regardless of the value of $b$, the system may undergo stochastic transitions between two stochastic states: a large-radius state and a small-radius state. When $b$ belongs to the deterministic birhythmicity region (Fig.~\ref{fig_DWP_samples}(b)), these states fluctuate around the two deterministic stable limit cycles. In this case, the deterministic unstable limit cycle naturally serves as the separatrix between them. When $b$ lies outside the birhythmicity region (Figs.~\ref{fig_DWP_samples}(a) and \ref{fig_DWP_samples}(c)), one state fluctuates around the single stable limit cycle, whereas the other is associated with the ghost of the annihilated limit cycle. The radius of this ghost state is approximated by the radius of the corresponding saddle-node limit cycle. In the deterministic system, depending on which side of the ghost the initial condition is located, trajectories either approach the ghost transiently or move away from it immediately. Thus, the ghost acts as an effective (\textit{de facto}) separatrix, making it a natural criterion for distinguishing between the large- and small-radius stochastic states.

\begin{figure}
\centering
\includegraphics[width=0.8\linewidth]{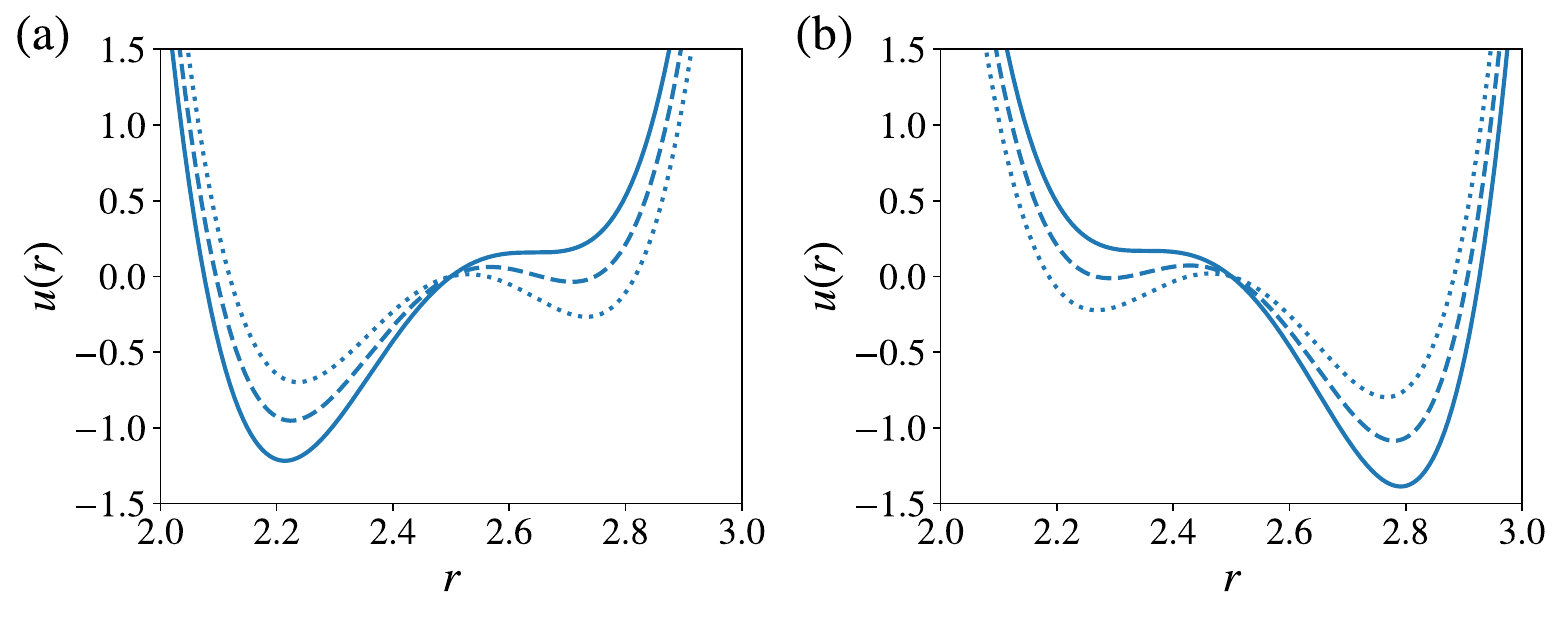}
\caption{Profiles of the scalar potential $u(r)$ around the saddle-node bifurcations. (a) Left saddle-node bifurcation for $b=-1.17$ (solid line), $b=-0.77$ (dashed line), and $b=-0.37$ (dotted line). (b) Right saddle-node bifurcation for $b=1.24$ (solid line), $b=0.84$ (dashed line), and $b=0.44$ (dotted line).}
\label{fig_DWP_u}
\end{figure}

Following the discussion above, we can define a probability $P_\text{LS}$ of observing the system evolving in the large-radius state at a given time. A numerical estimation of $P_\text{LS}$ over the $(b,\varepsilon)$ parameter plane, based on simulations of Eqs.~(\ref{eq_simple_model},\ref{eq_xi}), is shown in Fig.~\ref{fig_DWP_param_space}(a). For each pair $(b,\varepsilon)$, several simulations of duration $\Delta t=1000$ are performed using dispersed initial conditions, and the fraction of time steps for which $r(t)$ exceeds the separatrix radius is computed. For low noise levels ($\varepsilon<0.03$), $P_\text{LS}$ is nearly zero on the left side of the deterministic birhythmicity region, nearly one on the right side, and ranges between $0.36$ and $0.65$ inside it. The latter occurs because, under weak noise, the system rapidly approaches one or the other stable limit cycle depending on whether $r(0)$ lies below or above the separatrix, with little probability of crossing it during the subsequent evolution.

As the noise intensity increases, the region in which $P_\text{LS}$ is neither one nor zero becomes narrower. Inside the deterministic birhythmicity region, but close to either saddle-node bifurcation, the stable limit cycle created at the bifurcation is less stable than its counterpart. In terms of the scalar potential $u(r)$, this means that the well associated with its radius is shallower than that corresponding to the other attractor. This behavior is illustrated in Fig.~\ref{fig_DWP_u}. As the parameter $b$ crosses either saddle-node bifurcation, a new well appears in the profile of $u(r)$. Initially, this well is shallow, but it becomes progressively deeper as $b$ moves further into the deterministic birhythmicity region. Consequently, moderate noise can readily drive the system from the shallow well to the deeper one, whereas the reverse transition is much less likely. As a result, one state dominates the probability distribution, and both states are observed with comparable likelihood only near the center of the deterministic birhythmicity region. This defines a \textit{de facto} stochastic birhythmicity region. More specifically, this region is defined here as the set of points $(b,\varepsilon)$ for which both states have an observation probability of at least $2\%$, although this threshold is admittedly arbitrary.

The stochastic birhythmicity region reaches its minimum width around $\varepsilon\approx 4\times10^{-1}$. For higher noise intensities, the region broadens again, as fluctuations become sufficiently large to induce transitions between the two wells of $u(r)$ regardless of their relative depths. At $\varepsilon\approx 9\times10^{-1}$, the stochastic birhythmicity region becomes wider than its deterministic counterpart. In this regime, even outside the deterministic birhythmicity region, the system can switch between the stable limit cycle and the ghost of the annihilated cycle (see Fig.~\ref{fig_DWP_samples}). Overall, the simple model qualitatively reproduces the hourglass-shaped stochastic birhythmicity region observed in the stochastic Hindmarsh--Rose model \cite{Ortega_et_al_PRE_2025}. This suggests that the hourglass-shaped structure may be a robust feature of systems exhibiting birhythmicity bounded by two saddle-node bifurcations of limit cycles.

The probability $P_\text{LS}$ can also be theoretically estimated as the integral of the stationary distribution $p_s(r)$ from the separatrix radius discussed above to infinity,
\begin{equation}\label{eq_P_LS}
P_\text{LS} = \int_{r_\text{sep}}^{\infty} p_s(r)\,dr,
\end{equation}
This integral does not have a straightforward anti-derivative, but it is still relatively simple to compute numerically. The only point requiring special care is the choice of $r_\text{sep}$, the radius of the \textit{de facto} separatrix. As discussed above, in the deterministic monorhythmic regions this radius is identified with that of the annihilated saddle-node cycle, whereas in the deterministic birhythmic region it corresponds to the radius of the unstable limit cycle. The latter is given by the middle nonzero solution of $N(r^2)=b$. Although this equation is nontrivial, the unstable branch can be accurately fitted by a simple expression. Thus, $r_\text{sep}$ can be written as the following piecewise function of $b$,
\begin{equation}\label{eq_r_sep}
r_\text{sep} = \begin{cases}
               2.643, & b \leq -1.168, \\
               -0.0932\sqrt{b+1.168} + 0.0885\sqrt{1.238-b} + 2.5024, & -1.168 < b < 1.238, \\
               2.354, & b \geq 1.238.
               \end{cases}
\end{equation}
The numbers $-0.0932$, $0.0885$, and $2.5024$ are obtained by fitting the function to the unstable branch, with coefficient of determination $R^2=0.9998$. The other numbers are the coordinates of the saddle-node bifurcations (see Fig.~\ref{fig_DWP_det_bif_diag}).

The color map of the theoretical integral $P_\text{LS}$, Eq.~(\ref{eq_P_LS}), is shown in Fig.~\ref{fig_DWP_param_space}(b). The upper part of the hourglass-like structure is identical to that in Fig.~\ref{fig_DWP_param_space}(a). However, the lower part is absent. Instead, for $\varepsilon<0.2$, the probability $P_\text{LS}$ is nearly zero on the left side and nearly one on the right side, with an abrupt transition around $b=-0.021$, where the two wells of the potential $u(r)$ have equal depth. This discrepancy occurs because, in the lower central region of Fig.~\ref{fig_DWP_param_space}(a), the system has not yet reached stochastic equilibrium. Since the noise intensity is very low, there barely is any flow through the separatrix over the simulation time $\Delta t=1000$. The color map in Fig.~\ref{fig_DWP_param_space}(b) indicates that, under birhythmicity and very low noise, the system is much more likely to occupy the deeper potential well, even when the depth difference between the two wells is small. If the system is initially in the basin of attraction of the shallower well, it will first approach this well, but noise will eventually drive it into the deeper well, although this transition may require an extremely long time. This conclusion may seem counterintuitive, but it is consistent with Eq.~(\ref{eq_p_s}). For a potential $u(r)$ with two wells of similar depth, $-u(r)$ has two peaks of similar height. However, since $-u(r)$ is multiplied by $2/\varepsilon^2$, which is large for small $\varepsilon$, this small height difference is strongly amplified in the profile of $p_s(r)$. Therefore, the simulations in Fig.~\ref{fig_DWP_param_space}(a) would need to be run for much longer times in order to produce a map similar to that in Fig.~\ref{fig_DWP_param_space}(b). Presumably, the same mechanism occurs in the Hindmarsh--Rose model: the hourglass-shaped map obtained in Ref.~\cite{Ortega_et_al_PRE_2025} reflects the fact that the system requires very long times to reach stochastic equilibrium when the noise intensity is low.

\begin{figure}
\centering
\includegraphics[width=0.7\linewidth]{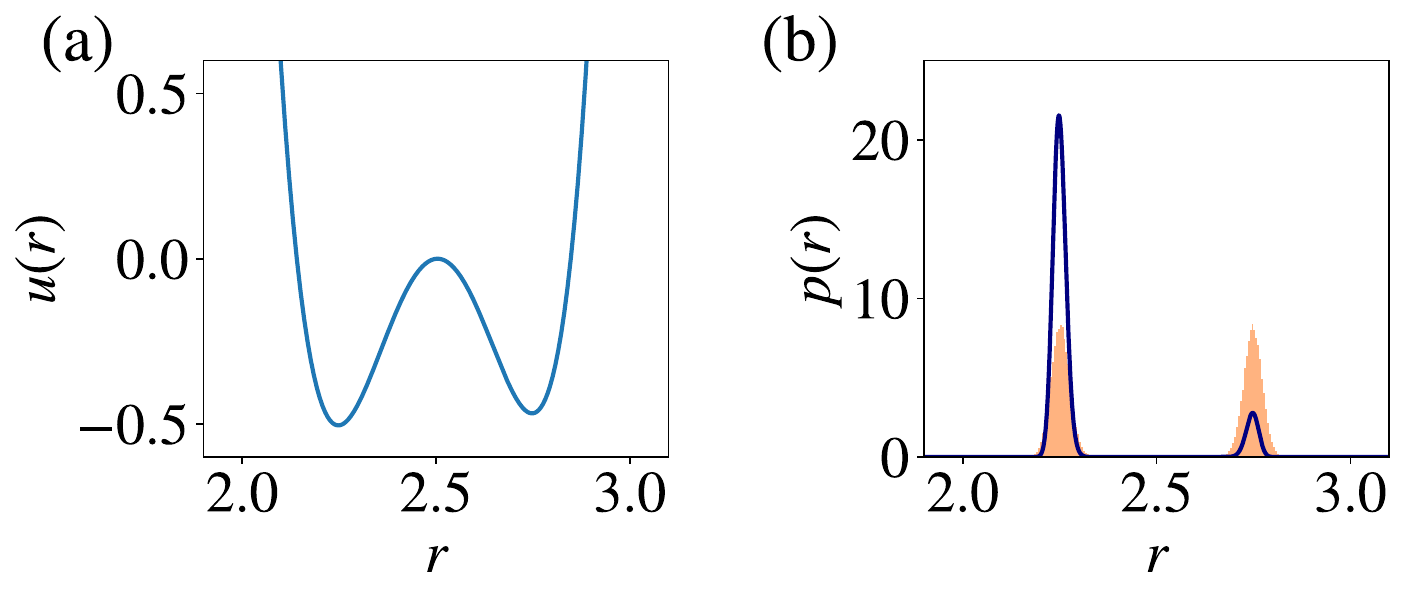}
\caption{(a) Profile of $u(r)$ for $b=-0.05$. (b) Normalized histogram of $r$ obtained from 100 simulations of Eqs.~(\ref{eq_simple_model},\ref{eq_xi}) for $b=-0.05$ and $\varepsilon=0.18$. The simulations have duration $\Delta t=10^{6}$ and use initial conditions uniformly distributed between $r=2$ and $r=3$. The histogram consists of 200 bins, shown in orange, and is constructed using only the last $0.05\%$ of the simulation data. The stationary probability distribution $p_s(r)$ is also shown as a solid blue curve.}
\label{fig_DWP_noneq}
\end{figure}

To further illustrate that, in the low-noise regime, the system requires an extremely long time to reach stochastic equilibrium, simulations of Eqs.~(\ref{eq_simple_model},\ref{eq_xi}) are performed with $b=-0.05$. For this value of $b$, the system is birhythmic and the potential $u(r)$ is nearly symmetric, with the two wells having almost the same depth, as shown in Fig.~\ref{fig_DWP_noneq}(a). The noise intensity is set to $\varepsilon=0.18$, and uniformly distributed initial conditions are used. These simulations have a very long duration, $\Delta t=10^{6}$, but only the last $0.05\%$ of the data is used to construct the histogram shown in Fig.~\ref{fig_DWP_noneq}(b). The histogram exhibits two sharp peaks around the deterministic attractors. This is intuitively expected, since the initial conditions are uniformly distributed, the attraction basins have similar sizes, and the noise intensity is small. However, the histogram does not match the stationary probability distribution $p_s(r)$. Indeed, $p_s(r)$ also has two sharp peaks around the deterministic attractors, but the left peak is much higher than the right one. This mismatch occurs because, even after such a long simulation time, the system has not yet reached stochastic equilibrium, which would require enough noise-induced transitions between attractors. Kramers' theory provides an order of magnitude of the expected escape time from a potential well, given by $\Delta t_\text{escape} \sim e^{\Delta u/\varepsilon^2}$, where $\Delta u$ is the well depth, i.e., the difference between the local maximum and the local minimum \cite{Kramers_Physica_1940}. For $\varepsilon=0.18$ and $\Delta u=0.5$ (see Fig.~\ref{fig_DWP_noneq}(a)), the escape time is of the order of $5\times 10^{6}$.

\subsection{Theoretical estimation of $P_\text{LS}$ in the low-noise regime}\label{subsec_low_noise}

As illustrated in Fig.~\ref{fig_DWP_param_space}, the colormap of the large state probability, $P_{\mathrm{LS}}$, estimated from numerical simulations, is well reproduced by the integral formula given in Eq.~\ref{eq_P_LS}, except in the deterministic birhythmicity region when the noise is low (namely, $\varepsilon<0.3$). Therefore, a theoretical expression that reproduces the simulational estimation of $P_{\mathrm{LS}}$ in this region of parameter space is yet to be found. To this end, we propose to carry out a procedure analogous to Kramers' escape rate problem \cite{Kramers_Physica_1940, Chandrasekhar_1943}.

We start from the Fokker--Planck equation, Eq.~\ref{eq_FP}. Owing to the axial symmetry of the system, we neglect the dependence on $\theta$. However, because we are now considering a regime in which the system has not yet reached stochastic equilibrium, the time dependence must be retained. This reduction is valid, for instance, when the initial probability distribution is axially symmetric. Thus, Eq.~\ref{eq_FP} yields
\begin{equation}\label{eq_FP_axial_sym}
\frac{\partial\rho}{\partial t} = \frac{1}{r} \frac{\partial}{\partial r} \bigg( r \Big( \frac{du}{dr}\rho + \frac{\varepsilon^2}{2} \frac{\partial\rho}{\partial r} \Big) \bigg).
\end{equation}

The probability density of the scalar coordinate $r$ is given by integrating $\rho(r,t)$ along the circumference of radius $r$ centered at the origin. Thus, $p(r,t) = 2\pi C(t) \rho(r,t)$, where $C(t)$ is a normalization coefficient. We also define an effective potential, $u_\text{eff}(r) = u(r) + \tfrac{1}{2} \varepsilon^2 \ln(r)$.  Multiplying Eq.~(\ref{eq_FP_axial_sym}) by $2\pi C(t)$ leads to
\begin{equation}\label{eq_FP_eff}
\frac{\partial p}{\partial t} = -\frac{\partial J_r}{\partial r},
\end{equation}
where $J_r$ is the radial probability flux given by
\begin{equation}\label{eq_J_r}
J_r = -\bigg( \frac{du_\text{eff}}{dr}p + \frac{\varepsilon^2}{2} \frac{\partial p}{\partial r} \bigg).
\end{equation}
Notice that Eq.~(\ref{eq_FP_eff}) is formally equivalent to the Fokker--Planck equation of a one-dimensional system evolving in the effective potential $u_\text{eff}$ and subjected to white noise of intensity $\varepsilon$.

In the deterministic birhythmicity region, the system has three equilibrium radii: the radius of the small stable limit cycle, the radius of the large stable limit cycle, and, between them, the radius of the unstable limit cycle. We denote these three radii by $r_{s1}$, $r_{s2}$, and $r_u$, respectively. As discussed above, $r_u$ acts as the separatrix between the basins of attraction of the two stable limit cycles. Therefore, the probability of observing the system in the large-radius state is obtained by integrating the probability distribution from $r_u$ to infinity,
\begin{equation}\label{eq_P_LS_noneq}
P_\text{LS} = \int_{r_u}^{\infty} p(r,t)\,dr.
\end{equation}
Subsequently, integrating Eq.~(\ref{eq_FP_eff}) in the same range leads to
\begin{equation}\label{eq_P_LS_diff}
\frac{dP_\text{LS}}{dt} = -J_r(r_u,t).
\end{equation}
We now aim to write Eq.~(\ref{eq_P_LS_diff}) as an ordinary differential equation for $P_\text{LS}$. To this end, we use the following identity satisfied by $J_r$:
\begin{equation}\label{eq_J_r_property}
\frac{\partial}{\partial r} \big( p \, e^{\frac{2}{\varepsilon^2}u_\text{eff}(r)} \big) = -\frac{2}{\varepsilon^2} J_r e^{\frac{2}{\varepsilon^2}u_\text{eff}(r)}.
\end{equation}

As discussed above, in the low-noise regime the system may require extremely long times to reach stochastic equilibrium. The evolution of the probability density is therefore slow, so that $\partial p/\partial t$ can be approximated as zero over the timescale considered. According to Eq.~(\ref{eq_FP_eff}), this implies that the radial probability flux $J_r$ depends only weakly on $r$. Hence, over the interval $[r_{s1},r_{s2}]$, $J_r$ can be approximated by its value at the separatrix, $J_r(r_u,t)$. Under this approximation, we integrate Eq.~(\ref{eq_J_r_property}) over $[r_{s1},r_{s2}]$,
\begin{equation}\label{eq_J_r_property_int}
p(r_{s2},t) e^{\frac{2}{\varepsilon^2}u_\text{eff}(r_{s2})} - p(r_{s1},t) e^{\frac{2}{\varepsilon^2}u_\text{eff}(r_{s1})} = -\frac{2}{\varepsilon^2} J_r(r_u,t) \int_{r_{s1}}^{r_{s2}} e^{\frac{2}{\varepsilon^2}u_\text{eff}(r)} \, dr.
\end{equation}

We now consider an initial probability density consistent with the initial conditions of the simulations, namely a uniform distribution over the interval $2\leq r\leq 3$. In the low-noise regime, the probability density rapidly concentrates around the deterministic attractors, forming two sharp peaks centered at $r_{s1}$ and $r_{s2}$ (see Fig.~\ref{fig_DWP_noneq}(b)). The areas under these peaks are initially determined by the amount of probability lying on each side of the separatrix $r_u$. Probability is then slowly transferred between the two attractors, so that the system evolves gradually toward stochastic equilibrium through changes in the peak heights. Since this transfer occurs on a much longer timescale than the local relaxation around each attractor, the density near each peak can be approximated by a quasi-equilibrium profile with a time-dependent amplitude. In particular, near the large-radius state $r_{s2}$, we approximate the probability density as
\begin{equation}
p(r,t) = p(r_{s2},t) e^{-\frac{2}{\varepsilon^2}(u_\text{eff}(r)-u_\text{eff}(r_{s2}))}.
\end{equation}
Next, we expand $u_\text{eff}(r)$ to second order around $r_{s2}$. Since $u_\text{eff}'(r_{s2})$ is approximated as zero, this gives
\begin{equation}\label{eq_p_Taylor}
p(r,t) = p(r_{s2},t) e^{-\frac{1}{\varepsilon^2}u_\text{eff}''(r_{s2})(r-r_{s2})^2}.
\end{equation}
Here, primes denote derivatives with respect to $r$. The approximation $u_\text{eff}'(r_{s2})\approx 0$ is justified because, for $\varepsilon<0.3$, the critical points of $u_\text{eff}(r)$ are very close to those of $u(r)$. Thus, the solutions of $u_\text{eff}'(r)=0$ are well approximated by the deterministic radii $r_{s1}$, $r_u$, and $r_{s2}$, which satisfy $u'(r)=0$.

We substitute Eq.~(\ref{eq_p_Taylor}) into Eq.~(\ref{eq_P_LS_noneq}). Assuming that the probability peak around $r_{s2}$ is narrow and well contained within the integration domain, the integral can be evaluated explicitly, yielding
\begin{equation}
P_\text{LS} = \sqrt{\frac{\pi \varepsilon^2}{u_\text{eff}''(r_{s2})}}\, p(r_{s2},t).
\end{equation}
Therefore, $p(r_{s2},t)$ can be expressed in terms of $P_\text{LS}$ as
\begin{equation}\label{eq_p_r_s2}
p(r_{s2},t) = \sqrt{\frac{u_\text{eff}''(r_{s2})}{\pi \varepsilon^2}}\, P_\text{LS}.
\end{equation}
Analogously, the value of the probability density at $r_{s1}$ can be expressed in terms of $P_\text{SS}$, the probability of observing the system in the small-radius state at a given time:
\begin{equation}\label{eq_p_r_s1}
p(r_{s1},t) = \sqrt{\frac{u_\text{eff}''(r_{s1})}{\pi \varepsilon^2}}\, P_\text{SS}.
\end{equation}
Since $r_{s1}$ and $r_{s2}$ are the radii of stable limit cycles, $u''(r_{s1})$ and $u''(r_{s2})$ are positive. In the low-noise regime, it is therefore reasonable to assume that $u_\text{eff}''(r_{s1})$ and $u_\text{eff}''(r_{s2})$ are also positive.

We now evaluate the integral on the right-hand side of Eq.~(\ref{eq_J_r_property_int}). Since $r_u$ is the radius of the unstable limit cycle, we have $u'(r_u)=0$ and $u''(r_u)<0$. In the low-noise regime, the integrand can be assumed to have a single dominant peak near $r=r_u$, well contained within the integration domain. Moreover, since $u_\text{eff}(r)$ remains close to $u(r)$ for sufficiently small $\varepsilon$, we can approximate $u_\text{eff}'(r_u)\approx 0$ and assume $u_\text{eff}''(r_u)<0$. Expanding $u_\text{eff}(r)$ to second order around $r_u$, the integral is approximated as
\begin{equation}\label{eq_integral}
\int_{r_{s1}}^{r_{s2}} e^{\frac{2}{\varepsilon^2}u_\text{eff}(r)} \, dr
\approx
\sqrt{-\frac{\pi \varepsilon^2}{u_\text{eff}''(r_u)}}\, e^{\frac{2}{\varepsilon^2}u_\text{eff}(r_u)}.
\end{equation}

Substituting Eqs.~(\ref{eq_p_r_s2}), (\ref{eq_p_r_s1}), and (\ref{eq_integral}) into Eq.~(\ref{eq_J_r_property_int}), we obtain an expression for the radial flux $J_r(r_u,t)$ in terms of $P_\text{LS}$ and $P_\text{SS}$:
\begin{equation}\label{eq_J_Rs}
J_r(r_u,t) = R_1 P_\text{SS} - R_2 P_\text{LS},
\end{equation}
where $R_1$ and $R_2$ are given by
\begin{align}
\label{eq_R1}
R_1 &= \frac{1}{2\pi} \sqrt{-u_\text{eff}''(r_{s1}) u_\text{eff}''(r_u)}\, e^{-\frac{2}{\varepsilon^2}(u_\text{eff}(r_u)-u_\text{eff}(r_{s1}))}, \\
\label{eq_R2}
R_2 &= \frac{1}{2\pi} \sqrt{-u_\text{eff}''(r_{s2}) u_\text{eff}''(r_u)}\, e^{-\frac{2}{\varepsilon^2}(u_\text{eff}(r_u)-u_\text{eff}(r_{s2}))}.
\end{align}

Substituting Eq.~(\ref{eq_J_Rs}) into Eq.~(\ref{eq_P_LS_diff}) and using $P_\text{SS}=1-P_\text{LS}$, we obtain the following differential equation for the time evolution of $P_\text{LS}$:
\begin{equation}
\frac{dP_\text{LS}}{dt} = -(R_1+R_2)P_\text{LS}+R_1.
\end{equation}

The solution of this equation is
\begin{equation}\label{eq_P_LS_sol}
P_\text{LS} = \frac{R_1}{R_1+R_2} + \bigg( P_\text{LS}^{~0} - \frac{R_1}{R_1+R_2} \bigg) e^{-(R_1+R_2)t},
\end{equation}
where $P_\text{LS}^{~0}$ denotes the initial value of $P_\text{LS}$. For an initial probability density uniformly distributed over the interval $2\leq r\leq 3$, this initial value is given by $P_\text{LS}^{~0}=(3-r_u)/(3-2)$. Equation~(\ref{eq_P_LS_sol}) depends on the three radii $r_{s1}$, $r_u$, and $r_{s2}$, which are the nonzero solutions of $N(r^2)=b$. Since this equation is not trivial to solve explicitly, these radii can be approximated by fitting simple functions to the corresponding branches, yielding
\begin{align}
r_{s1} &= -0.0938(1.238-b)^{0.4528} + 2.354, \\
r_u    &= -0.0932\sqrt{b+1.168} + 0.0885\sqrt{1.238-b} + 2.5024, \\
r_{s2} &=  0.0989(b+1.168)^{0.4592} + 2.643.
\end{align}
All three fits have a coefficient of determination of approximately $R^2=0.9998$.

\begin{figure}
\centering
\includegraphics[width=0.6\linewidth]{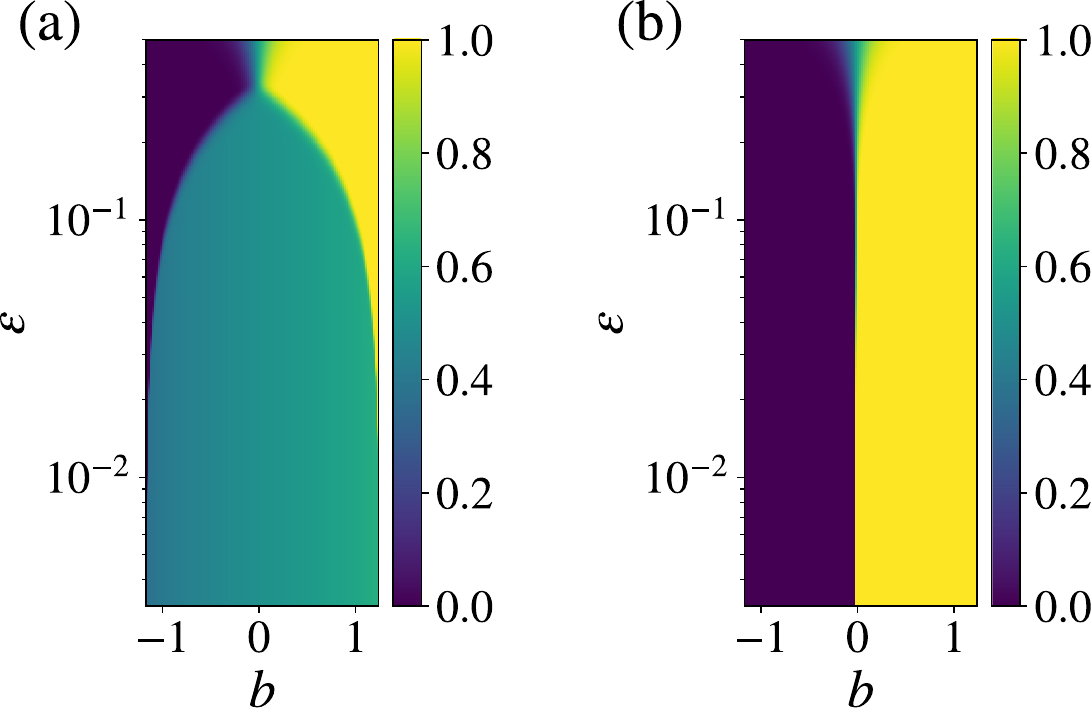}
\caption{Theoretical estimation of the probability $P_\text{LS}$ of observing the largest limit cycle, given by to Eq.~(\ref{eq_P_LS_sol}), as a function of $(b,\varepsilon)$, for (a) $t=1000$ and (b) $t=\infty$.}
\label{fig_Kramers}
\end{figure}

A colormap of the estimate of $P_\text{LS}$ obtained from Eq.~(\ref{eq_P_LS_sol}) is shown in Fig.~\ref{fig_Kramers}(a). Here, the time is set to $t=1000$, matching the duration of the numerical simulations. Equation~(\ref{eq_P_LS_sol}) reproduces the numerical estimate of $P_\text{LS}$ in the deterministic birhythmicity region for low noise, and the resulting map closely resembles that shown in Fig.~\ref{fig_DWP_param_space}(a). For very low noise, for example $\varepsilon\leq 0.01$, both $R_1$ and $R_2$ are negligible, so the exponential term in Eq.~(\ref{eq_P_LS_sol}) can be approximated by $1$. Therefore, $P_\text{LS}$ remains close to its initial value $P_\text{LS}^{~0}$, which, as discussed above, is equal to $3-r_u$. Along the deterministic birhythmicity interval, this quantity ranges from $0.36$ to $0.65$, in agreement with Figs.~\ref{fig_DWP_param_space}(a) and \ref{fig_Kramers}(a). For moderately low noise, namely $0.01<\varepsilon<0.3$, the exponential term in Eq.~(\ref{eq_P_LS_sol}) can no longer be approximated by $1$. In this case, both $P_\text{LS}^{~0}$ and $R_1/(R_1+R_2)$ contribute significantly to $P_\text{LS}$. In terms of accuracy, the difference between the value of $P_\text{LS}$ predicted by Eq.~(\ref{eq_P_LS_sol}) and that obtained from numerical simulations is below $0.05$ for most values of $(b,\varepsilon)$, and rarely exceeds $0.08$.

The estimate of $P_\text{LS}$ given by Eq.~(\ref{eq_P_LS_sol}) also captures the behavior expected at stochastic equilibrium, as illustrated in Fig.~\ref{fig_Kramers}(b). The result is very similar to that obtained from the stationary probability distribution in Fig.~\ref{fig_DWP_param_space}(b). As $t\to\infty$, the exponential term in Eq.~(\ref{eq_P_LS_sol}) vanishes, and $P_\text{LS}$ converges to $R_1/(R_1+R_2)$, which can be written as $1/(1+R_2/R_1)$. From Eqs.~(\ref{eq_R1}) and (\ref{eq_R2}), we obtain
\begin{equation}
\frac{R_2}{R_1} = \sqrt{\frac{u_\text{eff}''(r_{s2})}{u_\text{eff}''(r_{s1})}}\, e^{\frac{2}{\varepsilon^2}(u_\text{eff}(r_{s2})-u_\text{eff}(r_{s1}))}.
\end{equation}

As discussed above, there is a threshold value of the control parameter, approximately $b=-0.021$, at which the two attractors have the same potential depth, i.e., $u(r_{s1}) = u(r_{s2})$. The limiting value of $P_\text{LS}$ as $\varepsilon \to 0$ depends on whether $b$ lies below or above this threshold. If $b<-0.021$, then $u(r_{s1}) < u(r_{s2})$ (see Fig.~\ref{fig_DWP_u}(a)). Since $u_\text{eff}(r)\to u(r)$ as $\varepsilon\to 0$, the quantity $\frac{2}{\varepsilon^2}\left(u_\text{eff}(r_{s2})-u_\text{eff}(r_{s1})\right)$ diverges to $+\infty$. Consequently, $R_2/R_1$ also diverges, and therefore $1/(1+R_2/R_1)$ converges to $0$. On the other hand, if $b>-0.021$, then $u(r_{s1}) > u(r_{s2})$ (see Fig.~\ref{fig_DWP_u}(b)). In this case, $\frac{2}{\varepsilon^2}\left(u_\text{eff}(r_{s2})-u_\text{eff}(r_{s1})\right)$ diverges to $-\infty$ as $\varepsilon\to 0$. Hence, $R_2/R_1$ converges to $0$, and $1/(1+R_2/R_1)$ converges to $1$. This explains the step-like structure of the colormap shown in Fig.~\ref{fig_Kramers}(b).

\section{Double saddle-node bifurcation model with localized noise}\label{sec_simple_model_LN}

\begin{figure}
\centering
\includegraphics[width=\linewidth]{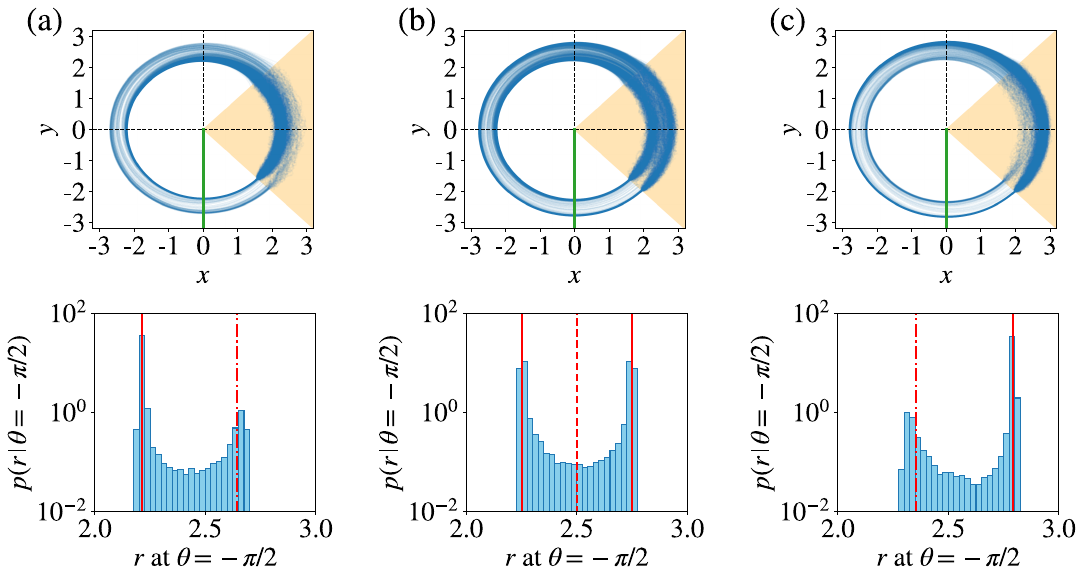}
\caption{Numerical simulations of Eqs.~(\ref{eq_simple_model_LN}, \ref{eq_xi_LN}, \ref{eq_chi}) for $\varepsilon=1.5$ and different values of $b$: (a) $b=-1.25$, (b) $b=0$, and (c) $b=1.25$. Results from 100 simulations with dispersed initial conditions are shown in phase space, together with the active region, shown in yellow. The radial distance $r$ is computed each time the trajectory crosses the line $\theta=-\pi/2$, shown as a solid green line, and the corresponding histogram is constructed. Each histogram also shows the radii of the stable and unstable deterministic limit cycles, indicated by solid and dashed red lines, respectively. In panels (a) and (c), the radius of the nearest saddle-node bifurcation is also shown as a dash-dotted red line.}
\label{fig_DWPLN_samples}
\end{figure}

Here, the model described in Sec.~\ref{sec_simple_model} is modified to emulate the asymmetric behavior of the Hindmarsh--Rose model, in which noise is likely to induce transitions between coexisting bursting attractors, or their ghosts, only in regions where the attractors lie close to one another. Rather than breaking the axial symmetry of the deterministic attractors in the simple model, we break the axial symmetry of the noise by restricting it to a specific angular sector, namely $-\pi/4<\theta<\pi/4$, corresponding to the yellow region in Fig.~\ref{fig_DWPLN_samples}. This region is relatively small compared with the rest of phase space, in analogy with the Hindmarsh--Rose model, where the coexisting limit-cycle trajectories remain close to one another in the bursting region, which is also relatively small compared with the rest of the trajectories. The model equation reads
\begin{equation}\label{eq_simple_model_LN}
\frac{d\mathbf{r}}{dt} = \mathbf{F}(\mathbf{r}) + \varepsilon\chi(\theta)\boldsymbol{\xi}(t),
\end{equation}
where $\mathbf{F}(\mathbf{r})$ is the deterministic field defined in Eqs.~(\ref{eq_F},\ref{eq_f},\ref{eq_N}). In this case, the angular velocity is fixed at $\omega=50$ so that more rotations around the origin are performed and more data can be collected in each simulation. This choice is nevertheless arbitrary. The bivariate, uncorrelated white-noise process $\boldsymbol{\xi}(t)$ is the same as in Eq.~(\ref{eq_xi}),
\begin{align}\label{eq_xi_LN}
  \begin{split}
    \langle \xi_i(t)          \rangle &= 0, \\
    \langle \xi_i(t) \xi_j(t')\rangle &= \delta_{ij}\delta(t'-t),
  \end{split}
\end{align}
Finally, $\chi(\theta)$ is a step function,
\begin{equation}\label{eq_chi}
  \chi(\theta) = \begin{cases}
    1    & \text{if } -\tfrac{\pi}{4} < \theta < \tfrac{\pi}{4},\\
    0.01 & \text{otherwise.}
  \end{cases}
\end{equation}
Thus, noise can trigger transitions between the circumferential stable limit cycles, or their ghosts, only within the region $\theta \in (-\pi/4, \pi/4)$. We will refer to this sector as the active region and the remainder of phase space as the quiescent region. Ultimately, the system approaches either an attractor or a ghost in the quiescent region, depending on its state upon leaving the active region. This behavior is illustrated in Fig.~\ref{fig_DWPLN_samples}, which shows numerical simulations of Eqs.~(\ref{eq_simple_model_LN}, \ref{eq_xi_LN}, \ref{eq_chi}). To determine which state the system approaches in the quiescent region, the radial distance $r$ is recorded each time the trajectory crosses the negative part of the $y$-axis, $\theta = -\pi/2$, and a histogram is constructed. Inside the deterministic birhythmicity region (Fig.~\ref{fig_DWPLN_samples}(b)), the system may converge to either stable limit cycle, as reflected by the two histogram peaks located at their respective radii. Outside, but close to, the deterministic birhythmicity region (Figs.~\ref{fig_DWPLN_samples}(a) and \ref{fig_DWPLN_samples}(c)), only one stable limit cycle exists; nevertheless, due to the ghost phenomenon, the system can also approach the precursor of the other cycle that is about to emerge at the saddle-node bifurcation. This is again reflected in the histograms, which show one peak at the radius of the stable cycle and another at the radius corresponding to the nearby saddle-node bifurcation.

\begin{figure}
\centering
\includegraphics[width=0.9\textwidth]{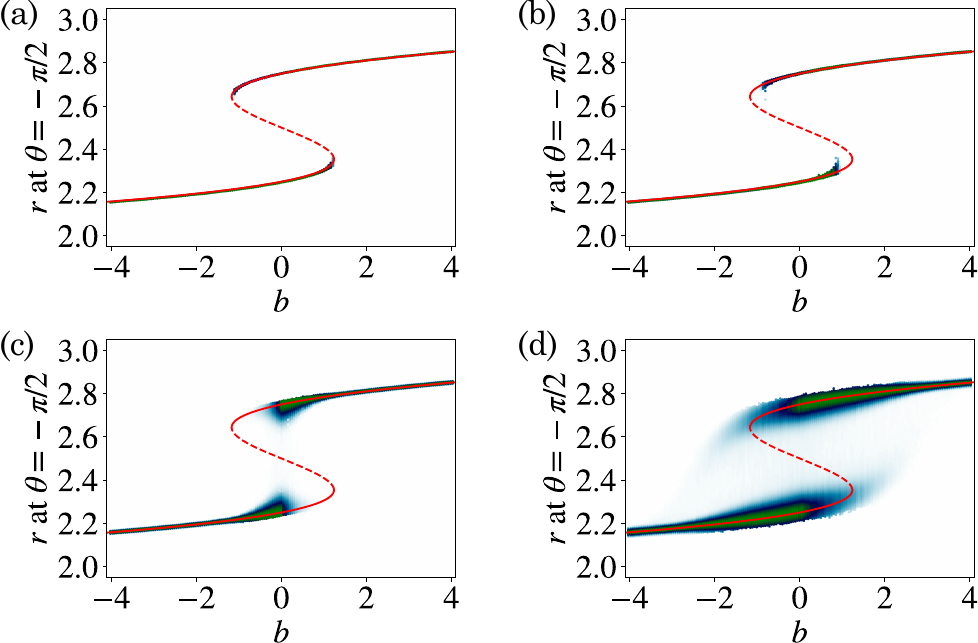}
\caption{Stochastic bifurcation diagrams of Eqs.~(\ref{eq_simple_model_LN},\ref{eq_xi_LN},\ref{eq_chi}) for different noise levels: (a) $\varepsilon=0.05$, (b) $\varepsilon=0.2$, (c) $\varepsilon=0.9$, and (d) $\varepsilon=2.5$. For each pair $(b,\varepsilon)$, 100 simulations are performed. Each time the trajectory crosses the lower part of the $y$-axis, corresponding to $\theta=-\pi/2$, the distance $r$ to the origin is recorded and plotted as a single point. The color of the points encodes their concentration, with blue indicating low density and green indicating high density. The deterministic stable and unstable limit-cycle branches are also shown as solid and dashed red lines, respectively.}
\label{fig_DWPLN_bif_diag}
\end{figure}

The statistical behavior of the model defined by Eqs.~(\ref{eq_simple_model_LN}, \ref{eq_xi_LN}, \ref{eq_chi}) is summarized in Fig.~\ref{fig_DWPLN_bif_diag}. For different values of $(b,\varepsilon)$, the radial distance $r$, measured at the end of the quiescent region, is represented using scatter plots. In the low-noise regime, as shown in Fig.~\ref{fig_DWPLN_bif_diag}(a), the points closely follow the deterministic bifurcation diagram. As the noise intensity increases (Figs.~\ref{fig_DWPLN_bif_diag}(b) and (c)), the scatter broadens, and the deterministic stable branches are no longer fully populated near the saddle-node bifurcations. This behavior arises because, in these parameter regions, the system is bistable, but one of the stable branches is significantly less stable than the other, corresponding to a shallower potential well (see Fig.~\ref{fig_DWP_u}). Consequently, trajectories tend to remain near the deeper well, while noise-induced transitions to the less stable branch become unlikely during the active phase.

This behavior closely parallels that observed in the symmetric version of the model discussed in Sec.~\ref{sec_simple_model} (see Fig.~\ref{fig_DWP_param_space}(a)). At still higher noise levels, as shown in Fig.~\ref{fig_DWPLN_bif_diag}(d), the scatter becomes more broadly distributed. However, both stable branches become populated again, since the noise is now sufficiently strong to induce transitions in both directions during the active phase. The ghost phenomenon also becomes evident in this regime, as the scatter extends beyond the deterministic stable branches and the saddle-node bifurcations, thereby producing an enlarged stochastic birhythmicity region.

\begin{figure}
\centering
\includegraphics[width=0.5\linewidth]{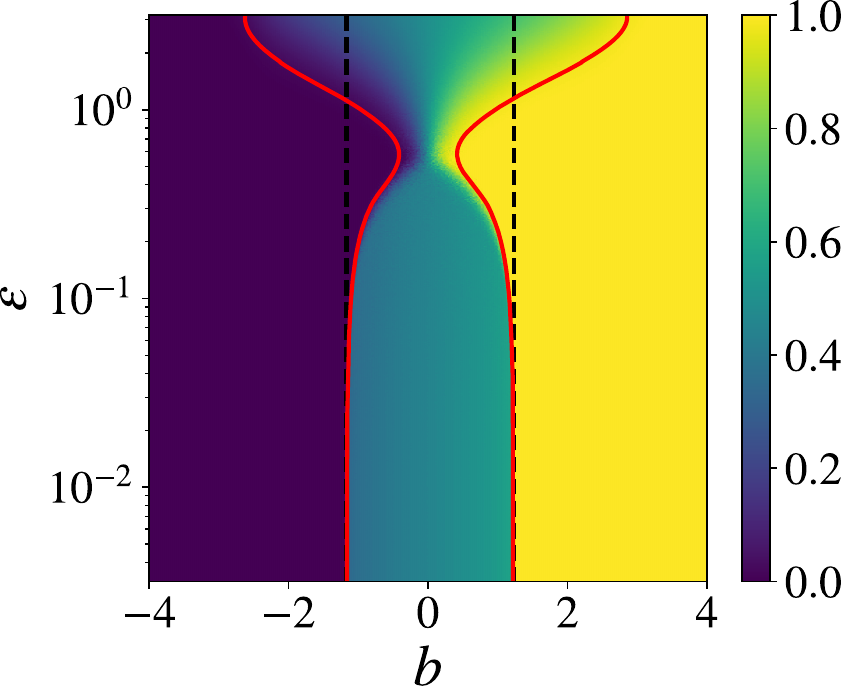}
\caption{Numerical estimation of the probability of observing the largest limit cycle, or its ghost, in the stochastic system, Eqs.~(\ref{eq_simple_model_LN}, \ref{eq_xi_LN}, \ref{eq_chi}), as a function of $(b,\varepsilon)$. For each parameter set, 100 simulations with uniformly distributed initial conditions and duration $\Delta t=1000$ are performed. After $t=500$, the values of $r$ are recorded each time the trajectory crosses the line $\theta=-\pi/2$, and the fraction of crossings for which $r$ lies above the mean radius of the stable limit cycles, or their ghosts, is computed. The dashed black lines indicate the boundaries of the deterministic birhythmicity region, while the solid red lines mark the estimated boundaries of the \textit{de facto} stochastic birhythmicity region, defined as the region in which both states occur with probability at least $2\%$.}
\label{fig_DWPLN_param_space}
\end{figure}

A more complete summary of the behavior of the model with localized noise over the $(b,\varepsilon)$ parameter plane is shown in Fig.~\ref{fig_DWPLN_param_space}. As discussed above, the system may approach two possible states during the quiescent region: either two stable limit cycles, or one stable limit cycle and a ghost, depending on whether $b$ lies inside or outside the birhythmicity region. For each pair $(b,\varepsilon)$, simulations of Eqs.~(\ref{eq_simple_model_LN}, \ref{eq_xi_LN}, \ref{eq_chi}) are performed over a time interval $\Delta t=1000$, using initial values of $r$ uniformly distributed between $r=2$ and $r=3$. The values of $r$ are recorded whenever the trajectory crosses the line $\theta=-\pi/2$, and the fraction of crossings for which $r$ is closer to the radius of the larger state is computed. For this calculation, the boundary between the two states is chosen as the mean of their radii, an admittedly arbitrary choice. The results qualitatively reproduce the hourglass-shaped profile observed in both the symmetric model of Sec.~\ref{sec_simple_model} and the Hindmarsh--Rose model.

As before, we define the \textit{de facto} stochastic birhythmicity region as the set of $(b,\varepsilon)$ values for which both states have a probability of observation of at least $2\%$ (although this cutoff is arbitrary). For $\varepsilon<10^{-1}$, the stochastic birhythmicity region coincides with its deterministic counterpart, since the noise is too weak to induce transitions and the outcome depends primarily on the initial conditions. As $\varepsilon$ increases, the stochastic birhythmicity region narrows, reaching a bottleneck around $\varepsilon\approx 6\times 10^{-1}$. For larger noise intensities, the region broadens again, becoming wider than the deterministic region at $\varepsilon\sim 10^{0}$, where fluctuations are strong enough to induce transitions between the stable limit cycles or their ghosts. Finally, outside the stochastic birhythmicity region, only the smaller state on the left, or the larger state on the right, is likely to be observed. It is important to note that, as in the homogeneous-noise case, the hourglass-like structure described above reflects finite-time behavior rather than stochastic equilibrium in the low-noise regime. Reaching stochastic equilibrium, and thus obtaining a cup-shaped map similar to that in Fig.~\ref{fig_DWP_param_space}(b), would require exceedingly long simulations. Moreover, such a map cannot be obtained from a simple radial integral such as Eq.~(\ref{eq_P_LS}), since the localized-noise model is not axially symmetric.

\section{Conclusions}\label{sec_conclusions}

In this paper, we have studied the effects of noise on birhythmic systems in which two limit cycles coexist within a region bounded by two saddle-node bifurcations. The Hindmarsh--Rose model, which exhibits multiple such regions where bursting periodic regimes coexist, is of particular interest. When noise is introduced into this model, noise-induced transitions between bursting regimes are observed even beyond the saddle-node bifurcations. In other words, birhythmic behavior may arise in parameter regions where it is not expected deterministically.

To provide an accessible description of this stochastic birhythmicity phenomenon, we first proposed a simple, axially symmetric model with homogeneous noise, in which two circumferential limit cycles coexist within a region delimited by two saddle-node bifurcations. The effect of noise intensity on birhythmicity was analyzed numerically by estimating, over the $(b,\varepsilon)$ parameter plane, the probability of observing the system near the largest limit cycle or its ghost. This probability displays an hourglass-shaped region of stochastic birhythmicity embedded between two regions of monorhythmicity. For sufficiently intense noise, this stochastic birhythmicity region extends beyond its deterministic counterpart. However, the hourglass-shaped profile is a finite-time effect: at low noise intensity, the system does not reach stochastic equilibrium within the simulation time, since the corresponding transition times become exceedingly long. A theoretical estimate based on the stationary probability distribution accounts for the high-noise regime, where stochastic equilibrium is reached within the simulation time, whereas the low-noise finite-time behavior is captured by an approximate approach inspired by Kramers' escape problem. This suggests that similar hourglass-shaped profiles may also arise in finite-time simulations of the Hindmarsh--Rose model and in other systems exhibiting birhythmicity bounded by saddle-node bifurcations of limit cycles. Since the phenomenon is captured by both a simple model and a complex neuronal model, it may also be observable in experimental setups with similar characteristics.

To account for the lack of symmetry in the Hindmarsh--Rose model, the simplified model was then modified by introducing non-homogeneous noise with enhanced intensity in a specific region of phase space. As a result, noise-induced transitions between the two limit cycles occur preferentially within this region, reproducing the fact that, in the Hindmarsh--Rose model, transitions between attractors take place mainly when the system evolves in the vicinity of both limit cycles. For this asymmetric model, a similar numerical analysis was performed, and an hourglass-shaped stochastic birhythmicity region was again observed. This supports the idea that such a structure is a recurrent feature of finite-time numerical simulations and experimental observations of systems displaying birhythmicity bounded by saddle-node bifurcations \cite{Wang_Yang_Duan_ERA_2023, TMY_CSF_2022}. Moreover, the existence of a noise intensity that minimizes the probability of transitions between the two limit cycles, corresponding to the bottleneck of the hourglass, appears to be a robust characteristic of this class of systems.

\section*{ACKNOWLEDGMENTS}
 This work has been supported by the Spanish State Research Agency (AEI) and the European Regional Development Fund (ERDF, EU) under Project No.~PID2023-148160NB-I00 (MCIN/AEI/10.13039/501100011033).

\section*{DATA AVAILABILITY STATEMENT}
The data that support the findings of this study are available from the corresponding author, upon reasonable request.

\end{document}